\documentclass{elsart}
\usepackage{graphicx}

\journal{Nuclear Instruments and Methods B}

\newcommand{\eps}{\varepsilon}
\newcommand {\bra}    {\langle}
\newcommand {\ket}    {\rangle}

\begin{document}

\begin{frontmatter}
\title{Positronium formation from valence and inner shells in noble gas atoms}

\author{L. J. M. Dunlop},
\ead{l.dunlop@qub.ac.uk}
\author{G.~ F.~Gribakin\corauthref{cor}}
\ead{g.gribakin@qub.ac.uk}
\corauth[cor]{Corresponding author}

\address{Department of Applied Mathematics and Theoretical Physics,
Queen's University Belfast, Belfast BT7 1NN, Northern Ireland, UK}
\begin{abstract}
When recent experimental positronium (Ps) formation cross sections in
noble gases have been compared with the most up-to date theoretical studies,
the agreement is qualitative, but not quantitative. In this paper we
re-examine this process and show that at low energies Ps formation must be
treated nonperturbatively. We also look at Ps formation with inner shell
electrons.  
\end{abstract}

\end{frontmatter}

\section{Introduction} 
Positronium (Ps) represents a bound state between a positron and an electron.
It is formed in positron-atom collisions,
\begin{equation}
A + e^+  \longrightarrow A^+ +{\rm Ps},
\end{equation}
when the positron energy, $\eps =k^2/2$, is above the Ps formation
threshold,
\begin{equation}
\eps > |\eps_n|-|E_{1s}|
\end{equation}
where $\eps_n$ is the energy of the bound electron atomic orbital $n$,
$E_{1s}\approx -6.8$ eV is the energy of the ground-state Ps, and $k$ is the
incident positron momentum (atomic units are used throughout).

Recently positronium formation in Ne, Ar, Kr and Xe has been determined
by two experimental groups \cite{la,ma}. The two sets of data are in fairly
good agreement, especially at lower energies. However, recent distorted-wave
Born approximation (DWBA) calculations \cite{gi} overestimate the cross
sections by a large factor, ranging from 1.6 in Ne to 3 in Xe, while the
overall energy dependence of the DWBA cross sections is in
reasonable accord with experiment. This is in contrast with earlier
coupled-static calculations \cite{mca}, which yield better magnitudes of the
cross section maxima, but disagree on the energy dependence.

In this paper we perform 1st-order and all-order calculations of Ps formation
from valence and subvalence subshells. Our consideration is restricted
to Ps formation in the ground-state. Noble gas atoms have tightly bound
electrons, making excited-state Ps formation much less probable (see, e.g.,
\cite{gi}). We argue that a structure observed at energies beyond the
main cross section maximum (described as a shoulder, or in some cases
seen as a secondary peak \cite{la,CCG83}) is most likely related to Ps
formation by the subvalence $ns$ electrons. We also consider Ps formation
from inner shells. It produces inner-shell vacancies and can be important
for positron-annihilation-induced Auger-electron spectroscopy \cite{oh}.

\section{1st-order approximation}\label{simapprox}

\subsection{Ps formation amplitude and cross section}\label{subsec:1st}

Using 1st-order many-body perturbation theory, and neglecting the interaction
between the outgoing Ps and residual ion, the Ps-formation amplitude can be
written as \cite{DFG96}
\begin{equation}\label{eq:ampl}
\langle\widetilde{\Psi}_{1s,{\bf K}}|V|n,\eps\rangle =
\int \widetilde{\Psi}^{*}_{1s,{\bf K}}({\bf r}_1,{\bf r}_2)
\left(-\frac{1}{|{\bf r}_1 -{\bf r}_2|}\right)
\psi_{n}({\bf r}_2)\varphi_{\eps}({\bf r}_1)d{\bf r}_1d{\bf r}_2,
\end{equation}
where $\varphi_{\eps}$ is the incident positron wavefunction,
$\psi_{n}$ is the Hartree-Fock wavefunction of the initial electron
state (``hole''), and $\widetilde \Psi _{1s,{\bf K}}$ is obtained from
the wavefunction of the ground-state Ps with momentum ${\bf K}$,
\begin{equation}\label{eq:Psi}
\Psi _{1s,{\bf K}}({\bf r}_1,{\bf r}_2)=
e^{i{\bf K}\cdot ({\bf r}_1+{\bf r}_2)/2} \phi_{1s}({{\bf r}_1-{\bf r}_2}),
\end{equation}
by orthogonalising it to all electron orbitals $n'$ occupied in the target
ground state,
\begin{equation}\label{eq:tilPsi}
\widetilde \Psi _{1s,{\bf K}}=\left(1-\sum _{n^\prime}|n^{\prime}\rangle
\langle n^{\prime}| \right) \Psi_{1s,{\bf K}}.
\end{equation}
The positron wavefunction is calculated in the field of the ground state atom
described in the Hartree-Fock approximation. The Ps center-of-mass motion is
described by a plane wave. The Ps formation cross section is found by
integration over the directions of ${\bf K}$,
\begin{equation}\label{eq:sigPs}
\sigma_{\rm Ps}=\frac{MK}{4\pi^2 k}\int
\bigl| \langle\widetilde{\Psi}_{1s,{\bf K}}|V|n,\eps\rangle \bigr|^2
d\Omega_{\bf K},
\end{equation}
where $M=2$ and $K=[2M(\eps - |\eps _n|+|E_{1s}|)]^{1/2}$ are
the Ps mass and momentum, and $\phi _\eps ({\bf r})\sim
e^{i{\bf k}\cdot {\bf r}}$ normalization is assumed. The approximation
(\ref{eq:ampl})--(\ref{eq:sigPs}) is equivalent to DWBA for a rearrangement
collision. 

A numerical calculation of the amplitude and cross section was performed
by expanding the Ps wavefunction in electron and positron spherical
harmonics with respect to the nucleus. Integration over the angular
variables was done analytically, while the radial integrals are
calculated numerically (see \cite{DFG96} for some details). To ensure accurate
positions of the Ps formation thresholds, experimental ionization energies
$|\eps _n|$, rather than the Hartree-Fock values, were used in the
calculations. Detailed below are the Ps-formation cross sections for neon,
argon, krypton and xenon, calculated using the 1st-order approximation
described above.

\subsection{Partial-wave contributions}

\begin{figure}[!ht]
\begin{center}
\includegraphics[width=8cm]{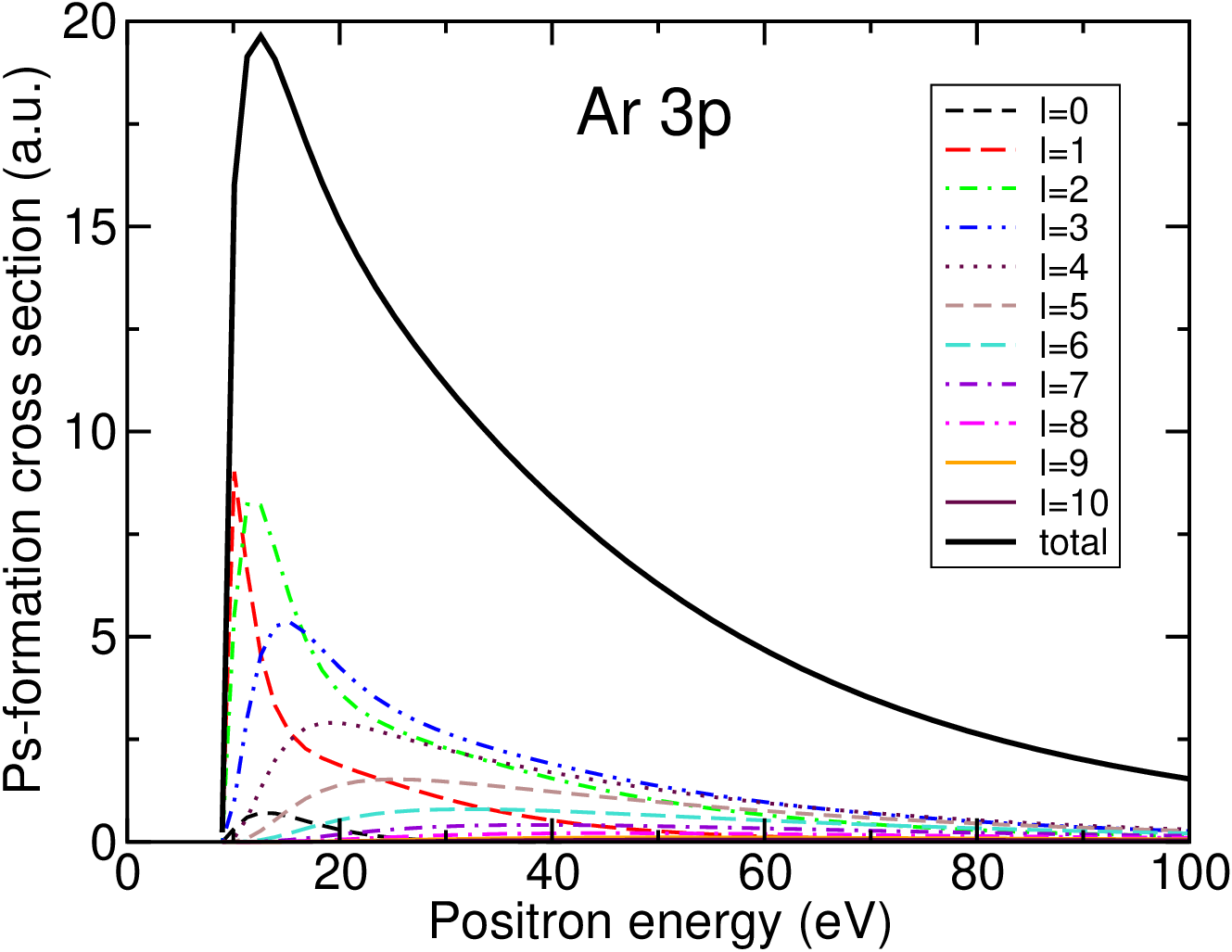}   
\caption{Partial wave contributions to the total Ps formation cross section for
the $3p$ subshell of argon. Various thin curves are the contributions
of $l=0$--10, while the solid thick curve is the total cross section.}
\label{arpart}
\end{center}
\end{figure}

The cross sections are found by summing over the positron partial waves
from $l=0$ to 10. Figure~\ref{arpart} shows the partial wave contributions for
the $3p$ subshell of argon. Note that the $p$, $d$, $f$ and $g$ waves have the
largest individual cross sections and make up most of the cross section peak.
The contributions of higher partial waves are suppressed by the centrifugal
barrier, preventing the close encounters which lead to Ps formation.
The small contribution of the $s$-wave is due to it being spherically
symmetric, making it harder for the positron to bind and move away with an
electron. This is true for all the noble gases.  The exceptionally small
$s$-wave contribution to Ps formation was noticed earlier for hydrogen and
helium (see, e.g., \cite{humberston,vanreeth1,vanreeth2}) and explained
by the hidden-crossing method \cite{ward}.

\subsection{Comparison with experiment}\label{subsec:comp}

\begin{figure}[ht]
\begin{center}
\begin{minipage}{6.8cm}
\includegraphics[height=5cm]{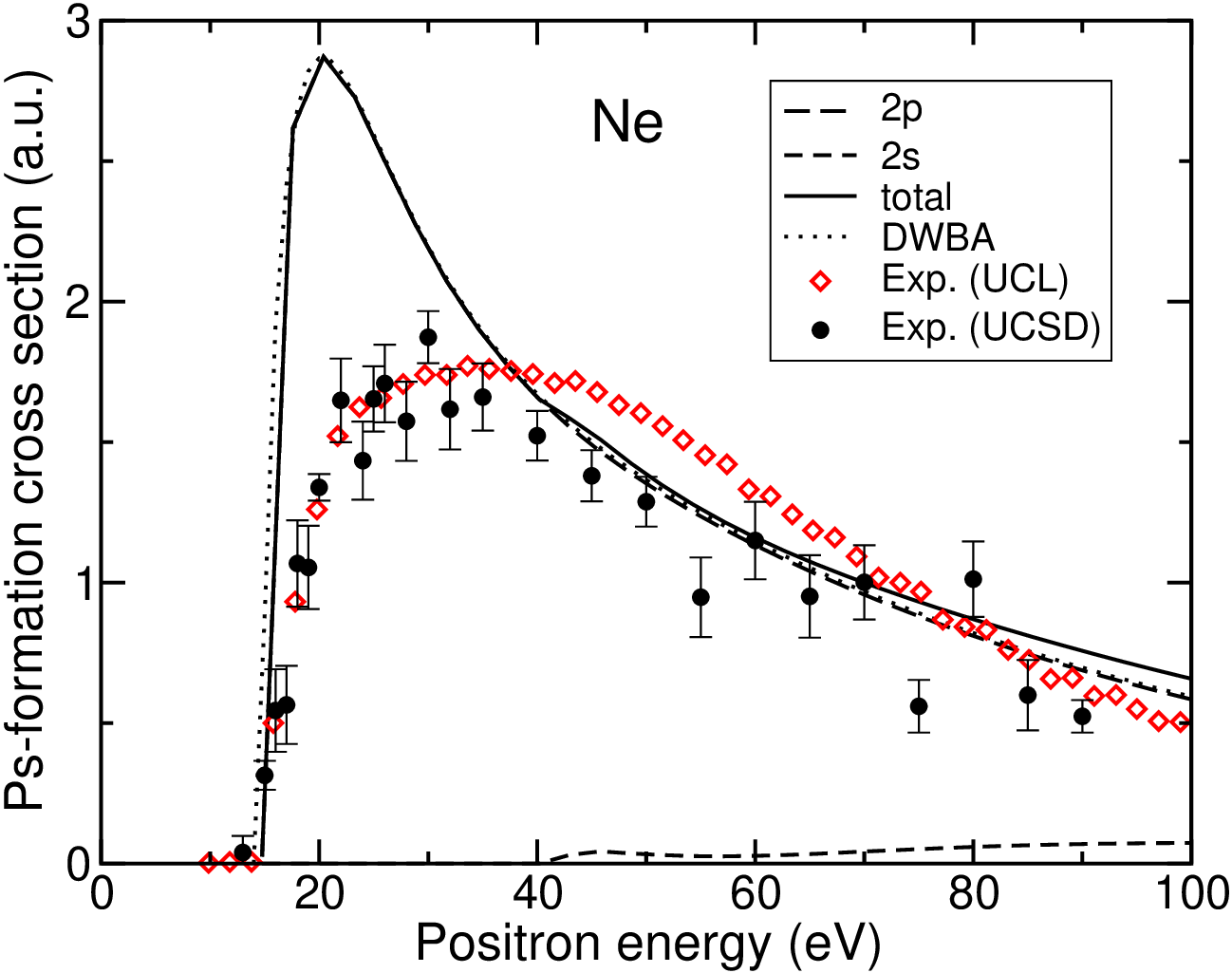}   
\end{minipage}
\begin{minipage}{6.8cm}
\includegraphics[height=5cm]{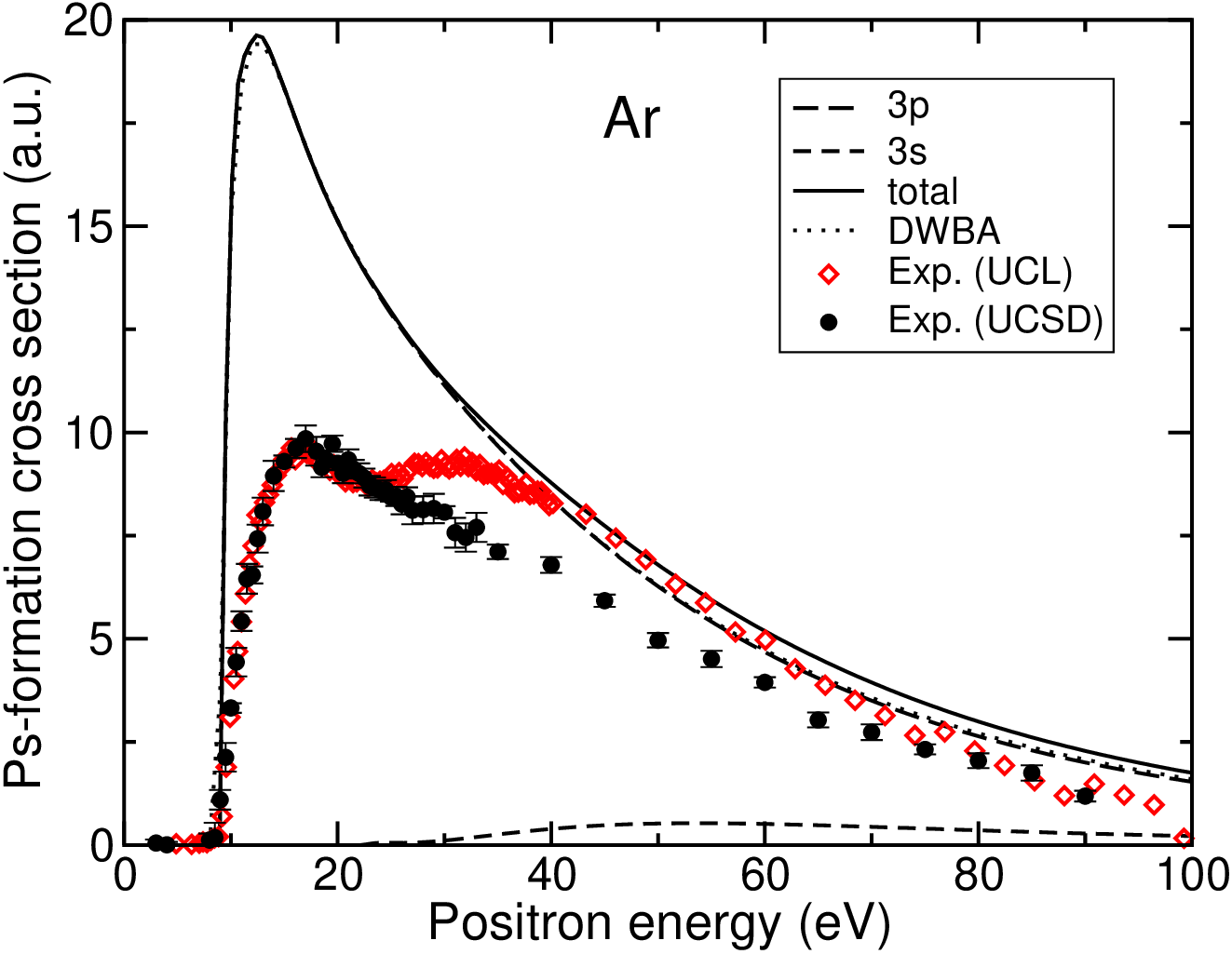}
\end{minipage}\\
\begin{minipage}{6.8cm}
\includegraphics[height=5cm]{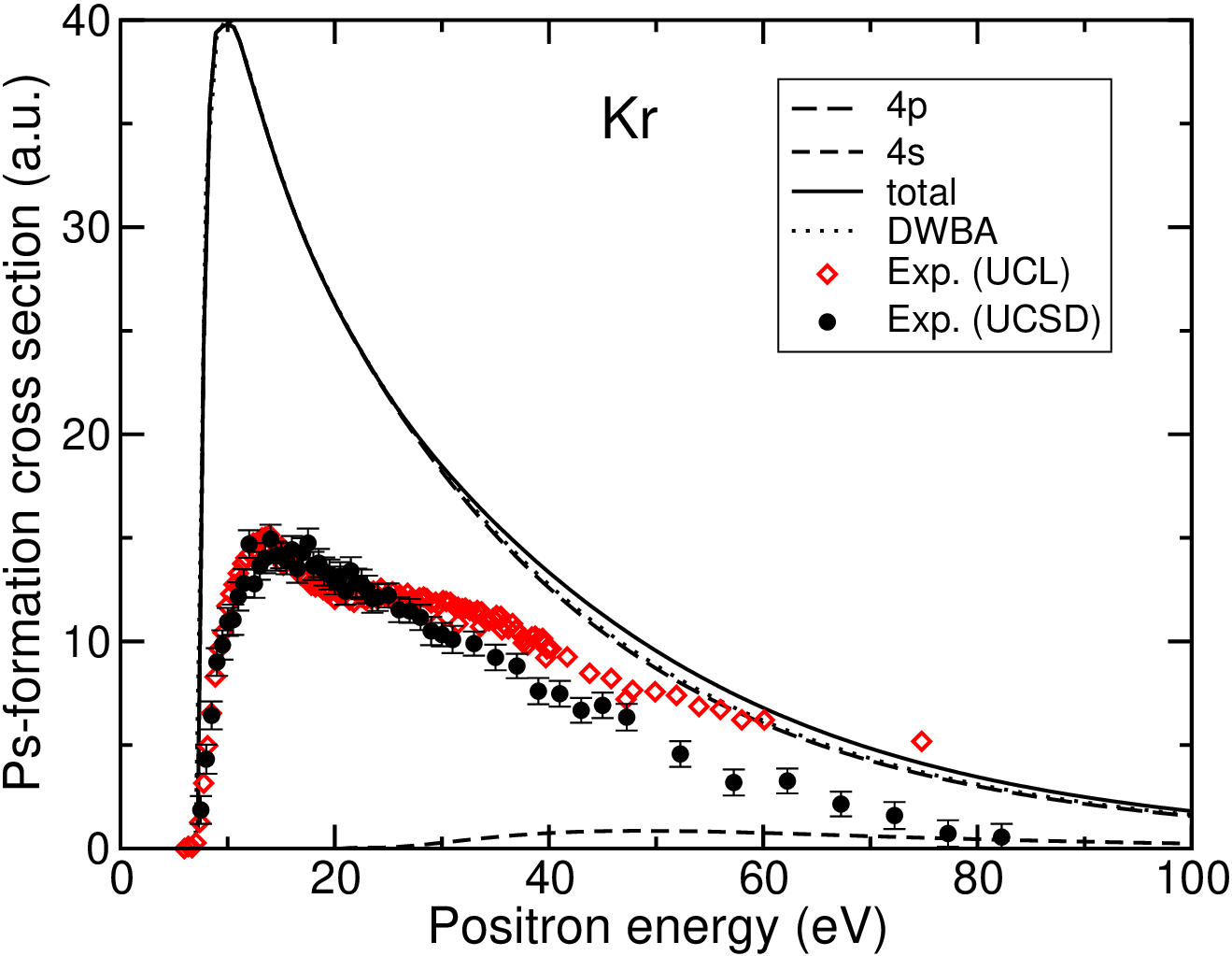}   
\end{minipage}
\begin{minipage}{6.8cm}
\includegraphics[height=5cm]{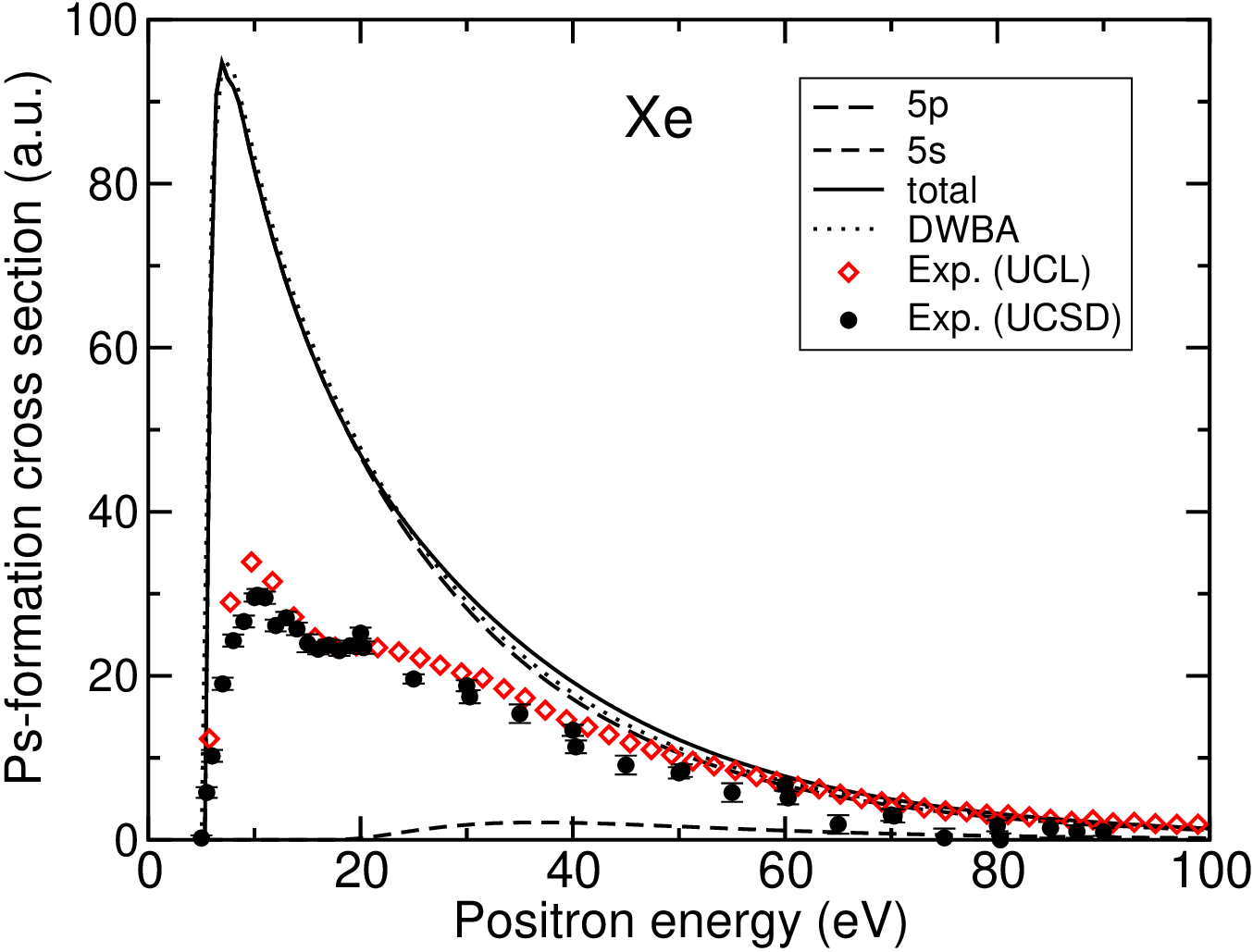}   
\end{minipage}
\caption{Comparison of the 1st-order Ps-formation cross sections for Ne, Ar,
Kr and Xe with experiment. Calculations:  long-dashed curve, contribution of
the $np$ subshell; short-dashed curve, contribution of the $ns$ subshell;
solid curve, total cross section; dotted curve, DWBA \cite{gi} for the $np$
subshell.  Experiment: solid circles, University of California
at San Diego (UCSD) \cite{ma}; open diamonds, University College London
(UCL) \cite{la}.}
\label{smprs}
\end{center}
\end{figure}

Figure~\ref{smprs} shows the Ps-formation cross sections for the valence
$np$ and subvalence $ns$ orbitals together with their sum, for
Ne, Ar, Kr and Xe. The present results for the $np$ subshell practically
coincide with the Ps($1s$) formation cross section from DWBA \cite{gi}.
The calculations are compared with the experimental data obtained with a
cold, trap-based positron beam \cite{ma}, and with the cross section
found by subtracting the direct ionization cross section from the total
ionization cross section \cite{la}. Moving from Ne to Xe, the calculations
increasingly overestimate the measured cross section near the maximum.

For Ne, Ar and Xe experiment and theory converge at higher energies,
while in Kr the discrepancy persists. Such convergence should be
expected from a theory point of view. Indeed, at higher positron energies
the dominant contribution to the amplitude (\ref{eq:ampl}) and cross section
(\ref{eq:sigPs}) comes from higher partial waves, for which the plane-wave
description of the Ps motion is more accurate. At the same time, the
contributions of individual partial waves to the amplitude become small.
This means that higher-order corrections neglected by the 1st-order theory
may not be important (see below). Thus, we cannot offer an explanation for
the divergence between theory and experiment in Kr.

\subsection{Inner-shell Ps formation}
\begin{figure}[!ht]
\begin{center}
\begin{minipage}{6.8cm}
\includegraphics*[height=5cm]{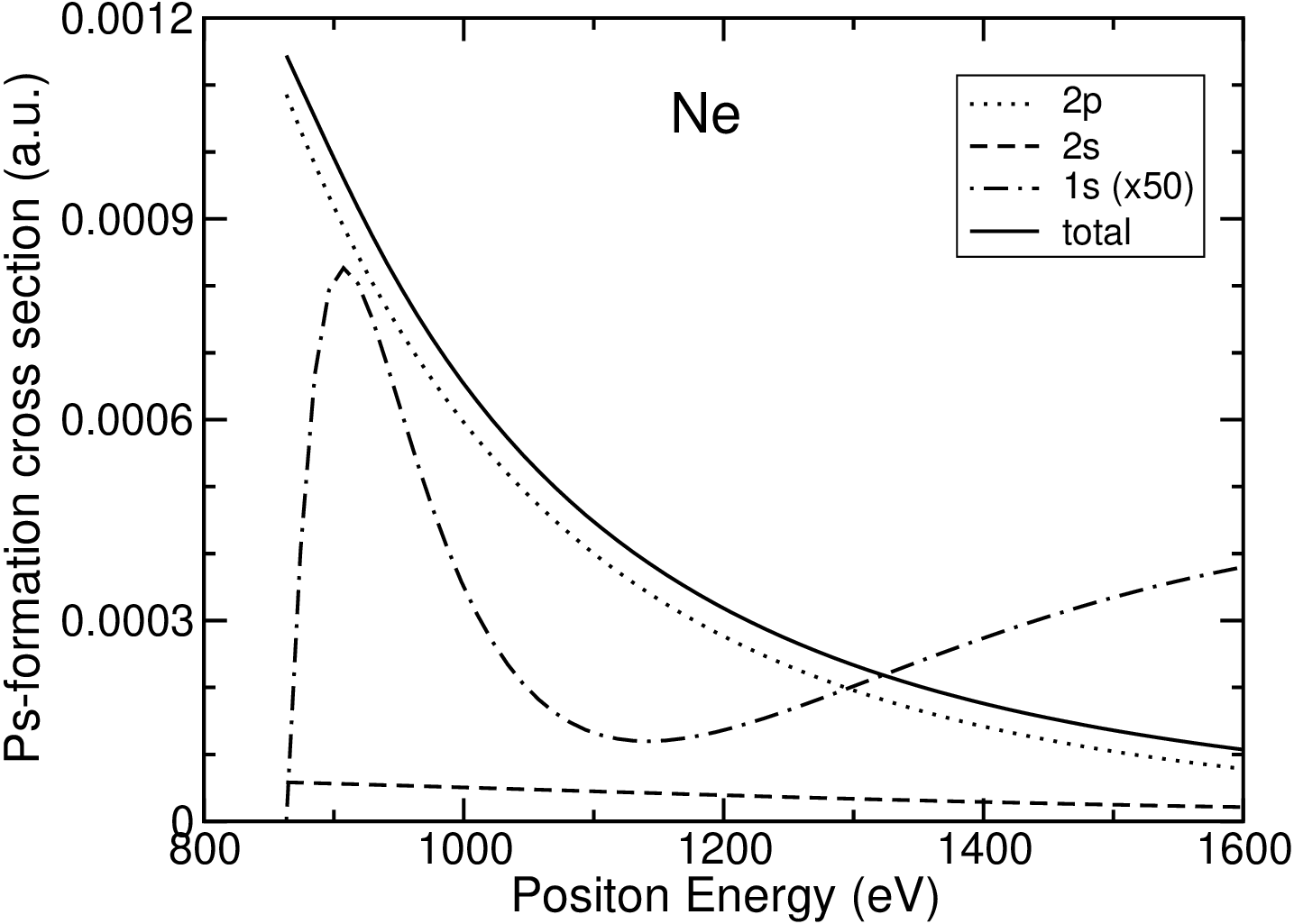}
\end{minipage}
\begin{minipage}{6.8cm}
\includegraphics*[height=5cm]{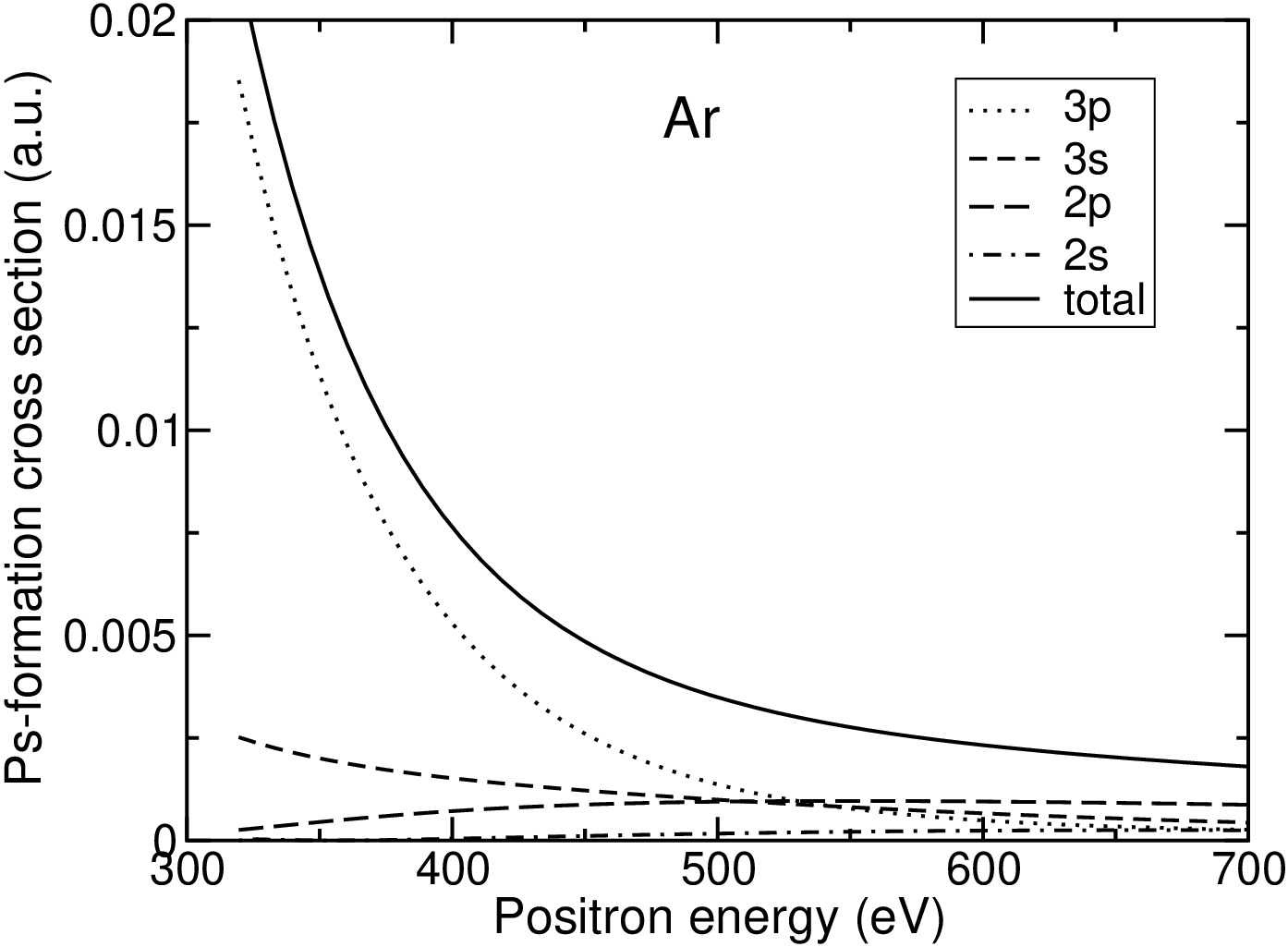}
\end{minipage}\\
\begin{minipage}{6.8cm}
\includegraphics*[height=5cm]{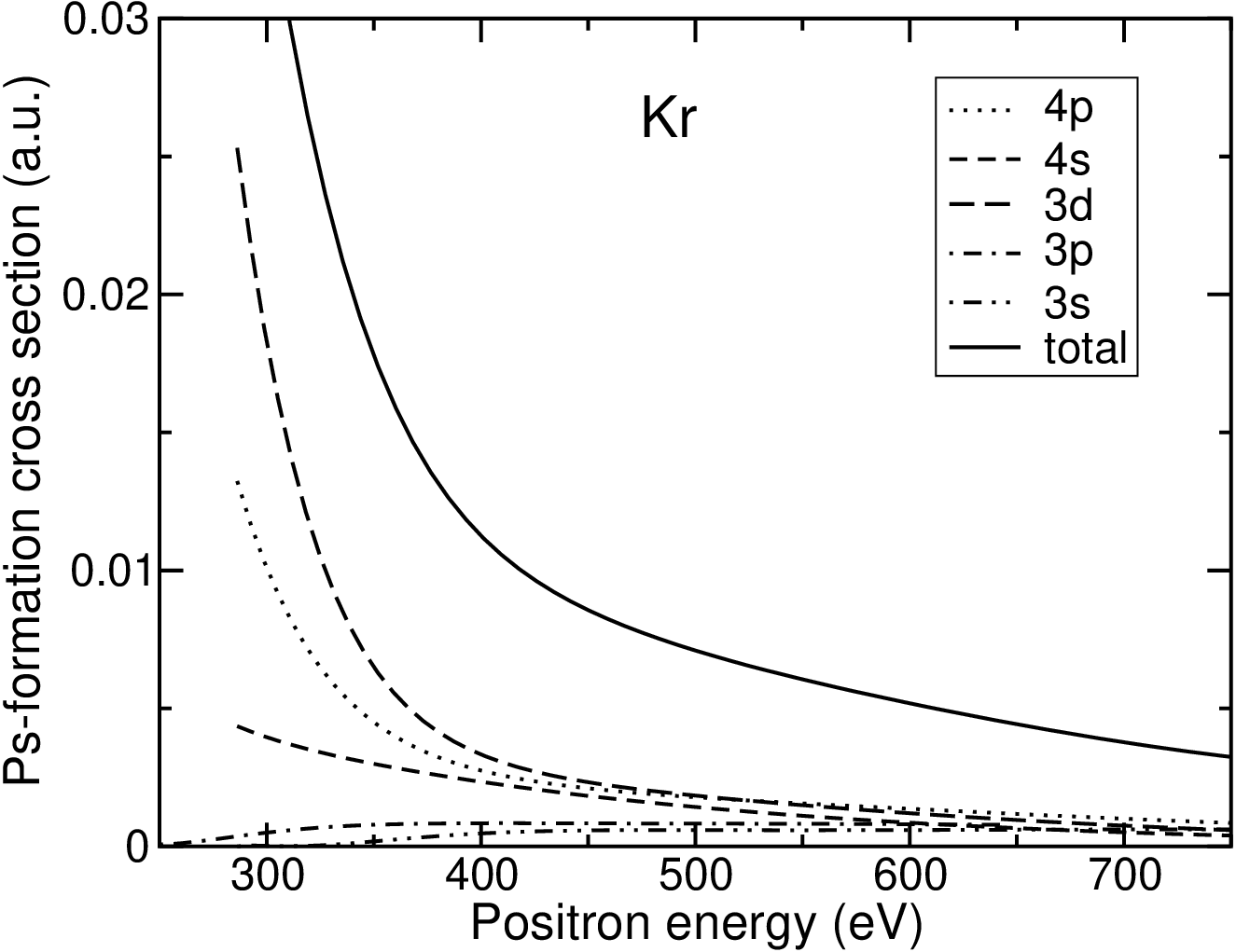}
\end{minipage}
\begin{minipage}{6.8cm}
\includegraphics*[height=5cm]{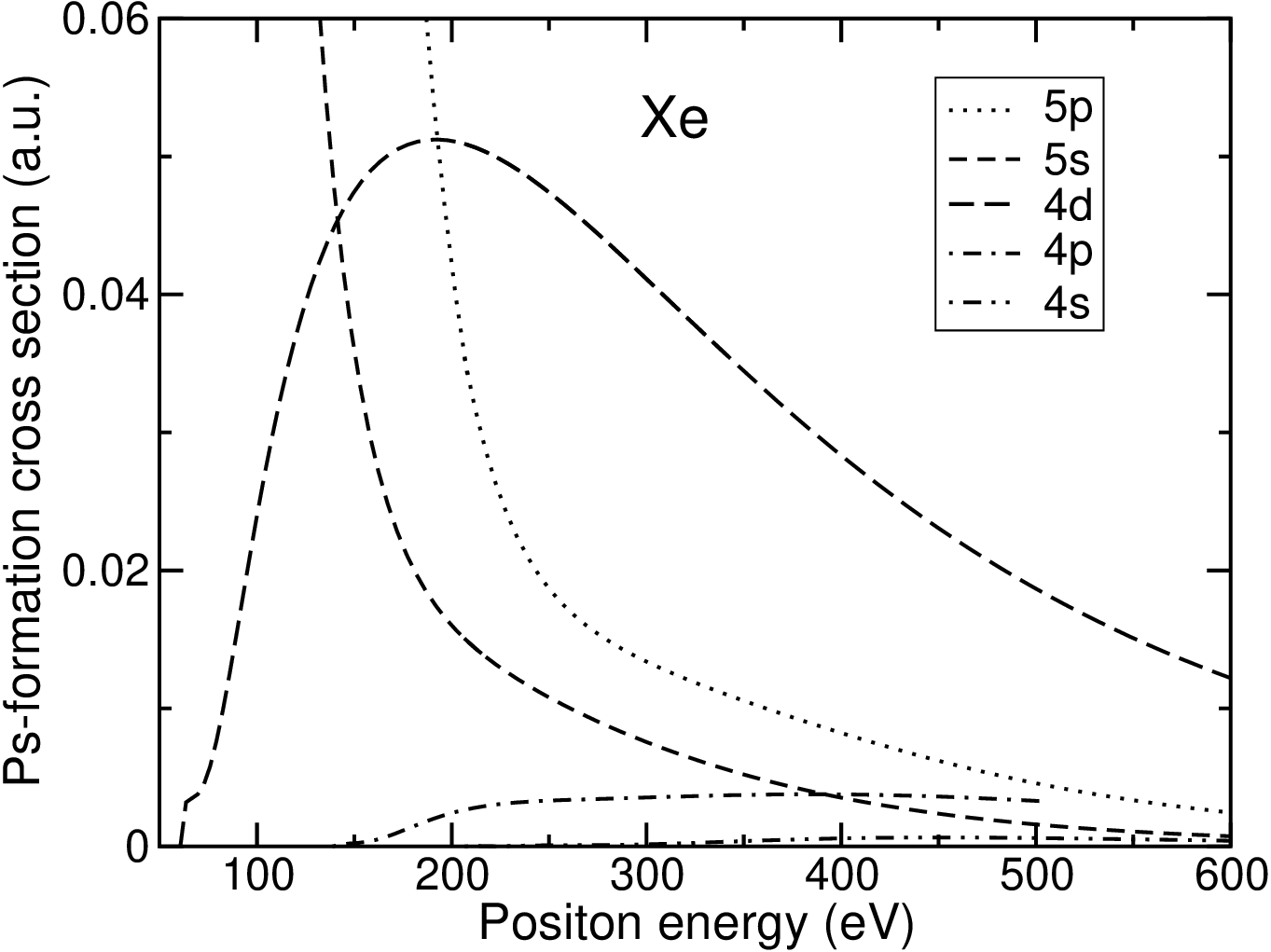}
\end{minipage}
\caption{Ps-formation cross sections for inner and valence subshells of
Ne, Ar, Kr and Xe.}
\label{engh}
\end{center}
\end{figure}

Ps-formation thresholds for the inner shells lie at much higher energies,
e.g., at 242 and 320~eV for the $2p$ and $2s$ orbitals in Ar. As a result,
the incident positron wavefunction oscillates rapidly, reducing the magnitude
of the amplitude (\ref{eq:ampl}). Ps formation from inner shells is
additionally suppressed by the positron repulsion from the nucleus.
The Ps-formation cross sections for Ne, Ar, Kr and Xe are shown in
Figure~\ref{engh}. It can be noted that significantly higher energies are
required to produce positronium for the lighter noble gases (with the more
tightly bound electrons).

In general, the Ps-formation cross sections for the inner-shell electrons are
quite small. Thus, in neon, the valence shell contribution still dominates,
even at positron energies of 1600~eV. In argon, it can be seen that up to
500~eV, the valence contribution dominates the cross section. However,
above this energy all subshells contribute approximately equally to the
total cross section, and  as the energy increases further, the $2p$ subshell
contributes the most. In krypton, the $3d$ subshell contributes most to 
the Ps-formation cross section, as this subshell is relatively far away 
from the nucleus. This is also true for xenon, where the $4d$ subshell
dominates the cross section. Note also that in Kr and especially Xe the
inner-shell cross sections are much larger than those of Ne and even Ar,
with their Ps formation threshold values nearly ten times smaller than
that of $1s$ in Ne.

\section{Ps formation: nonperturbative approach}

\subsection{Check of unitarity}\label{unit}

\begin{figure}[!ht]
\begin{center}
\begin{minipage}{6.8cm}
\includegraphics*[width=6.8cm]{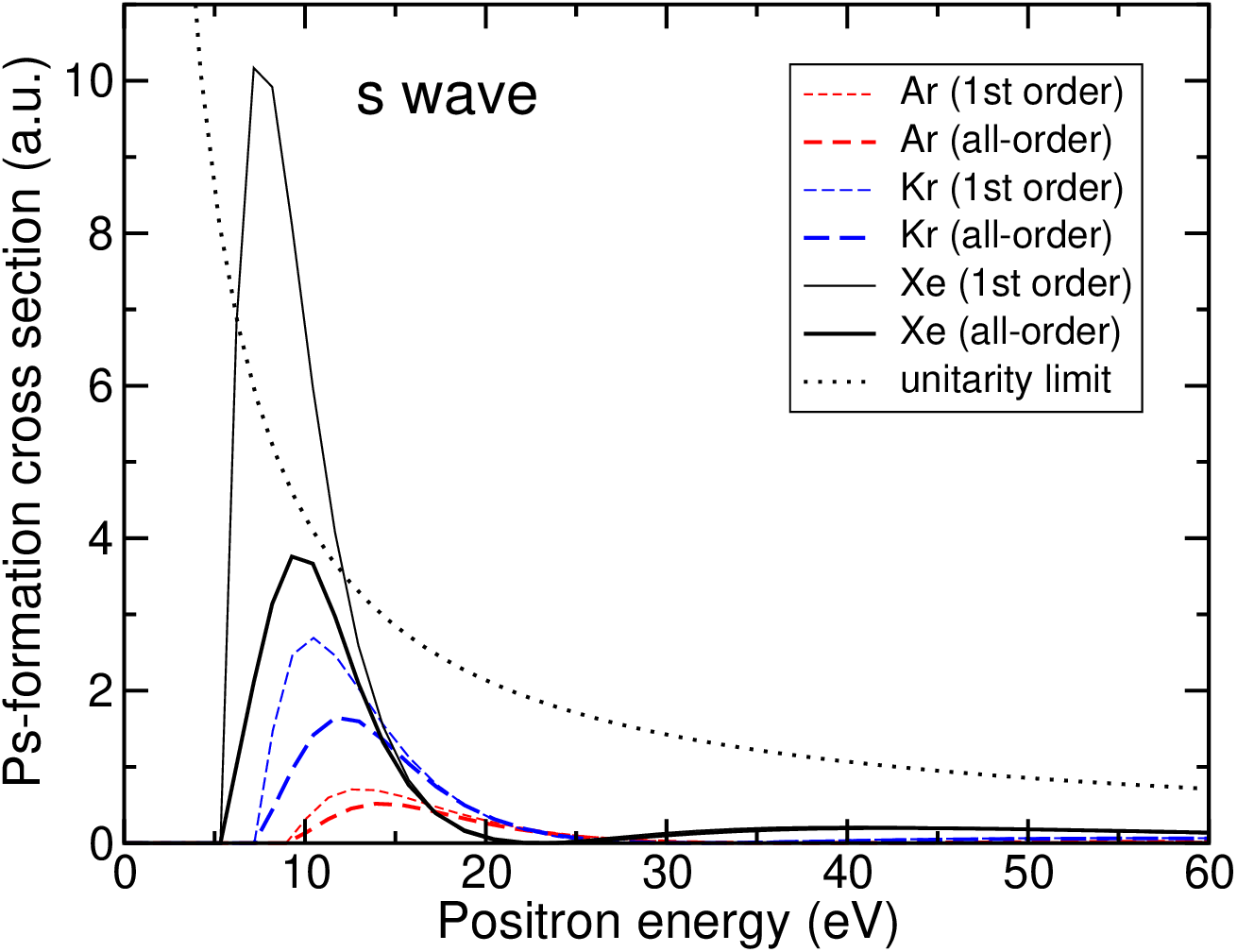}
\end{minipage}
\begin{minipage}{6.8cm}
\includegraphics*[width=6.8cm]{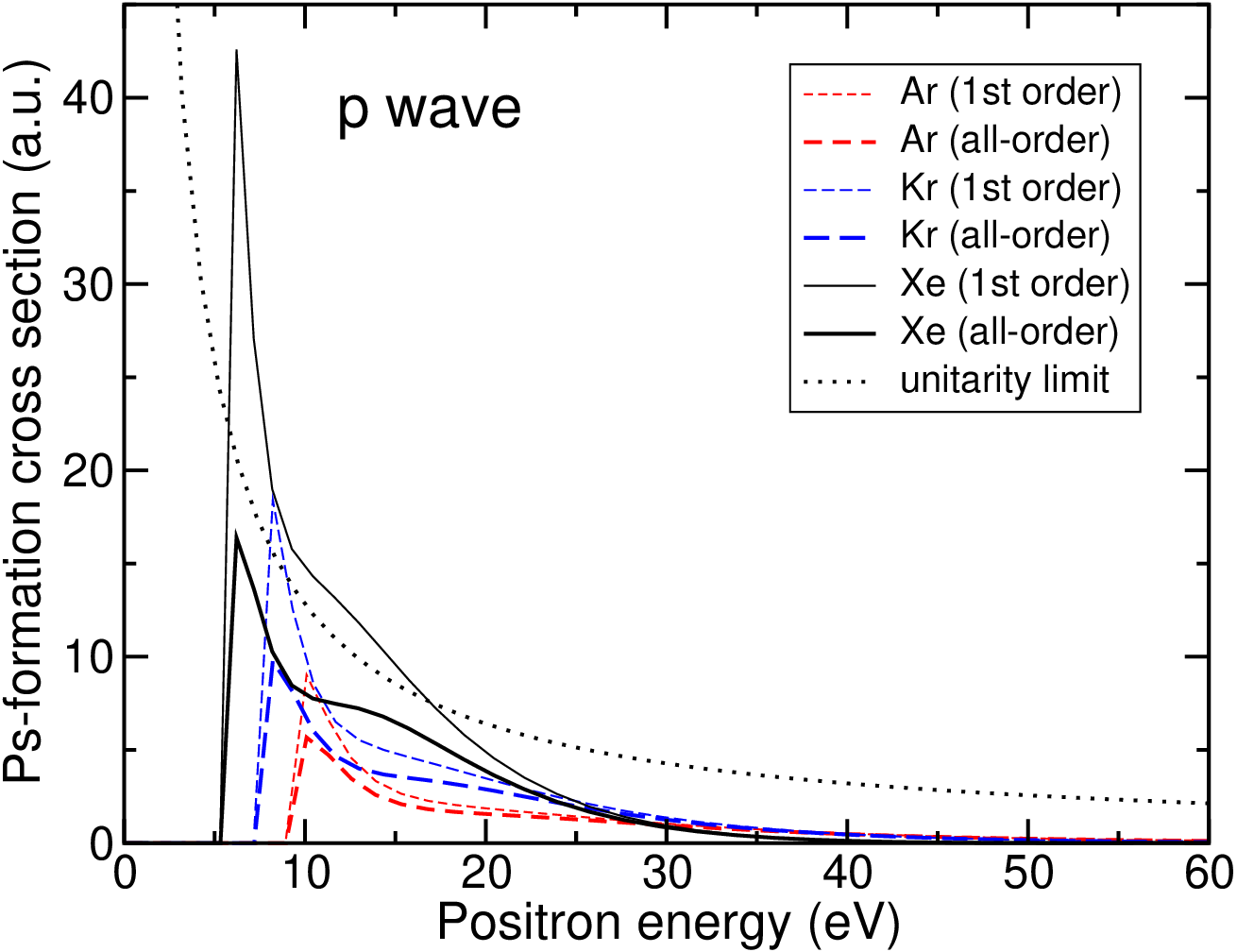}
\end{minipage}
\caption{Comparison of the $s$- and $p$-wave partial Ps formation cross
sections for Ar, Kr and Xe with the unitarity limit. Thin curves, 1st-order
approximation; thick curves, all-order approximation; dotted curves,
unitarity limit, $\pi(2l+1)/k^2$.}
\label{unit2}
\end{center}
\end{figure}

As seen from Figure \ref{smprs}, the Ps formation cross sections increase
dramatically from Ne to Xe. This increase is matched by a growing
discrepancy between the 1st-order results and experiment. This suggests
that as the cross sections become larger, the lowest-order perturbation
theory treatment becomes increasingly inaccurate.
Indeed, it turns out that in the 1st-order approximation, equations
(\ref{eq:ampl})--(\ref{eq:sigPs}), the lower partial-wave contributions
($l=0$--3) which dominate near the cross section maximum, become close to
and even violate (for Kr and Xe) the unitarity limit for inelastic processes,
$\sigma_{{\rm Ps}}^{(l)}\leq \pi (2l+1)/k^2$. A comparison with the unitarity
limit is presented in Figure \ref{unit2} for the $s$ and $p$ partial waves in
Ar, Kr and Xe.

Physically, unitarity
ensures that the amount of Ps formed in positron-atom collisions cannot be
greater than the number of positrons going in. Any inelastic cross section
above this limit is physically impossible. The fact that our 1st-order
results (and the analogous DWBA cross sections of Ref. \cite{gi})
are close to, or exceed the unitarity limit means that Ps formation is a
strong process which cannot be treated perturbatively. Hence, in addition to
the  1st-order Ps-formation amplitude (\ref{eq:ampl}), one must include
higher-order contributions which account for the decay of $A^+\mbox{Ps}$
back into the $e^++A$ channel, followed by Ps formation, etc. In other words,
the effect of Ps formation on the incident positron must be taken into account.
We do this by means of an all-order approach outlined in the next section.

\subsection{All-order approximation}\label{subsec:all}

The effect of Ps formation on the positron can be described by the Ps-formation
contribution to the positron-atom correlation potential, defined
by its matrix elements \cite{DFG96},
\begin{equation}\label{eq:Sigma}
\langle \eps '|\Sigma ^{({\rm Ps})}_E|\eps \rangle =
\int \frac{\langle \eps ',n|V|\widetilde{\Psi}_{1s,{\bf K}}\rangle
\langle\widetilde{\Psi}_{1s,{\bf K}}|V|n,\eps\rangle }
{E+\eps _n-E_{1s}-K^2/4+i0}\,\frac{d^3K}{(2\pi )^3},
\end{equation}
where $\langle\widetilde{\Psi}_{1s,{\bf K}}|V|n,\eps\rangle $
is the amplitude (\ref{eq:ampl}), $E_{1s}+K^2/4$ in the denominator
is the Ps energy, and the integral is over all Ps momenta ${\bf K}$.
Note that in this section we use positron states with a given angular
momentum, i.e., spherical waves,
$\varphi _\eps ({\bf r})=r^{-1} Y_{lm}(\Omega )P_{\eps l}(r)$,
with the radial wavefunctions normalized by
$P_{\eps l}(r)\sim (\pi k)^{-1/2}\sin (kr -
\frac12 \pi l +\delta _l^{\rm HF})$, where $\delta _l^{\rm HF}$ is the
positron phaseshift in the static field of the Hartree-Fock atom.
In this case the correlation potential (\ref{eq:Sigma}) is determined 
separately for each positron partial wave.

Below the Ps-formation threshold
$\bra\eps '|\Sigma^{(\rm{Ps})}_E|\eps \ket$
is real. Above the Ps-formation threshold, for $E>|\eps_n|-|E_{1s}|$,
the correlation potential acquires an imaginary part. This gives rise to
``absorption'' of the positron flux, which is being redirected into the
Ps formation channel. In fact, the 1st-order Ps formation cross
section (\ref{eq:sigPs}) is proportional to the imaginary part
of $\bra\eps |\Sigma^{(\rm{Ps})}_\eps |\eps\ket$,
\begin{equation}\label{eq:ImSig}
\mbox{Im} \bra\eps|\Sigma^{(\rm{Ps})}_\eps |\eps\ket
=-\pi \frac{MK}{(2\pi)^3}\int \left|\bra \tilde{\Psi}_{1s,{\bf K}}|V|
n \eps \ket\right|^2d\Omega_{\bf K},
\end{equation}
where $K=2(\eps - |\eps _n|+|E_{1s}|)^{1/2}$. Because
of the different normalization of the positron states adopted
in Secs. \ref{simapprox} and \ref{subsec:all}, caution is required when
using equations (\ref{eq:sigPs}) and (\ref{eq:ImSig}) to relate
$\sigma _{\rm Ps}$ to
${\rm Im} \bra\eps|\Sigma^{(\rm{Ps})}_\eps |\eps\ket $
(such relation is derived below).

A nonperturbative (``all-order'') calculation of Ps formation is done by
solving an integral equation for the matrix elements of
$\tilde{\Sigma} _E$ (see, e.g., \cite{ACC82}),
\begin{equation}\label{eq:redsig}
\bra\eps^{\prime}|\tilde{\Sigma} _E|\eps\ket
=\bra\eps^{\prime}|\Sigma^{(\rm{Ps})}_E|\eps\ket+
\int\frac{\bra\eps ^{\prime}|\tilde{\Sigma}_E|\eps ''\ket
\bra\eps^{\prime\prime}|\Sigma^{(\rm{Ps})}_E|\eps\ket}
{E-\eps^{\prime\prime}}
d\eps^{\prime\prime}.
\label{matele}
\end{equation}
The positron scattering phaseshift is then obtained as
\begin{equation}\label{eq:phsh}
\delta_l\equiv \delta_l^{\prime}+i\delta_l^{\prime\prime}=
\delta_l^{\rm{HF}}+\Delta\delta_l
\end{equation}
with
\begin{equation}\label{eq:Deldel}
\tan \Delta\delta_l=
-\pi\bra\eps|\tilde{\Sigma}_\eps |\eps\ket ,
\end{equation}
where $\Delta\delta_l$ is the additional phaseshift due to the correlation
potential. For energies $E$ above the Ps formation threshold,
the correlation potential, $\Sigma^{ \rm (Ps)}_E$, is complex,
and the phaseshift has a nonzero imaginary part, $\delta_l^{\prime\prime}>0$. 
The Ps-formation cross section is then obtained from
$\delta_l^{\prime\prime}$ by summing over the partial waves,
\begin{equation}\label{cssall}
\sigma_{\rm{Ps}}=\frac{\pi}{k^2}
\sum_{l=0}^\infty (2l+1)(1-e^{-4\delta_l''}).
\end{equation}

If we assume that Ps formation is a {\it weak} process, then 
$\bra\eps|\Sigma^{(\rm{Ps})}_E|\eps\ket$ is small, and we have
$\bra\eps|\tilde{\Sigma}_E|\eps\ket\approx
\bra\eps|\Sigma^{(\rm{Ps})}_E|\eps\ket$, 
$\Delta\delta _l\approx-\pi\bra\eps|\tilde{\Sigma}_\eps |\eps\ket$, and 
$\delta_l''\approx -\pi\mbox{Im}\bra\eps|\Sigma^{(\rm{Ps})}_\eps |\eps\ket $. 
Using $\delta_l^{\prime\prime}\ll 1$, we then have from equation
(\ref{cssall}):
\begin{equation}\label{eq:sig_Sig}
\sigma_{\rm Ps}^{(l)}\approx 
-\frac{4\pi^2}{k^2}(2l+1){\rm Im}\bra\eps|\Sigma^{(\rm{Ps})}|\eps\ket.
\end{equation}
The right-hand side of this equation is equivalent to the 1st-order
approximation examined earlier in this paper.

\subsection{Total Ps-formation cross sections}

\begin{figure}[!ht]
\begin{center}
\begin{minipage}{6.8cm}
\includegraphics*[height=5cm]{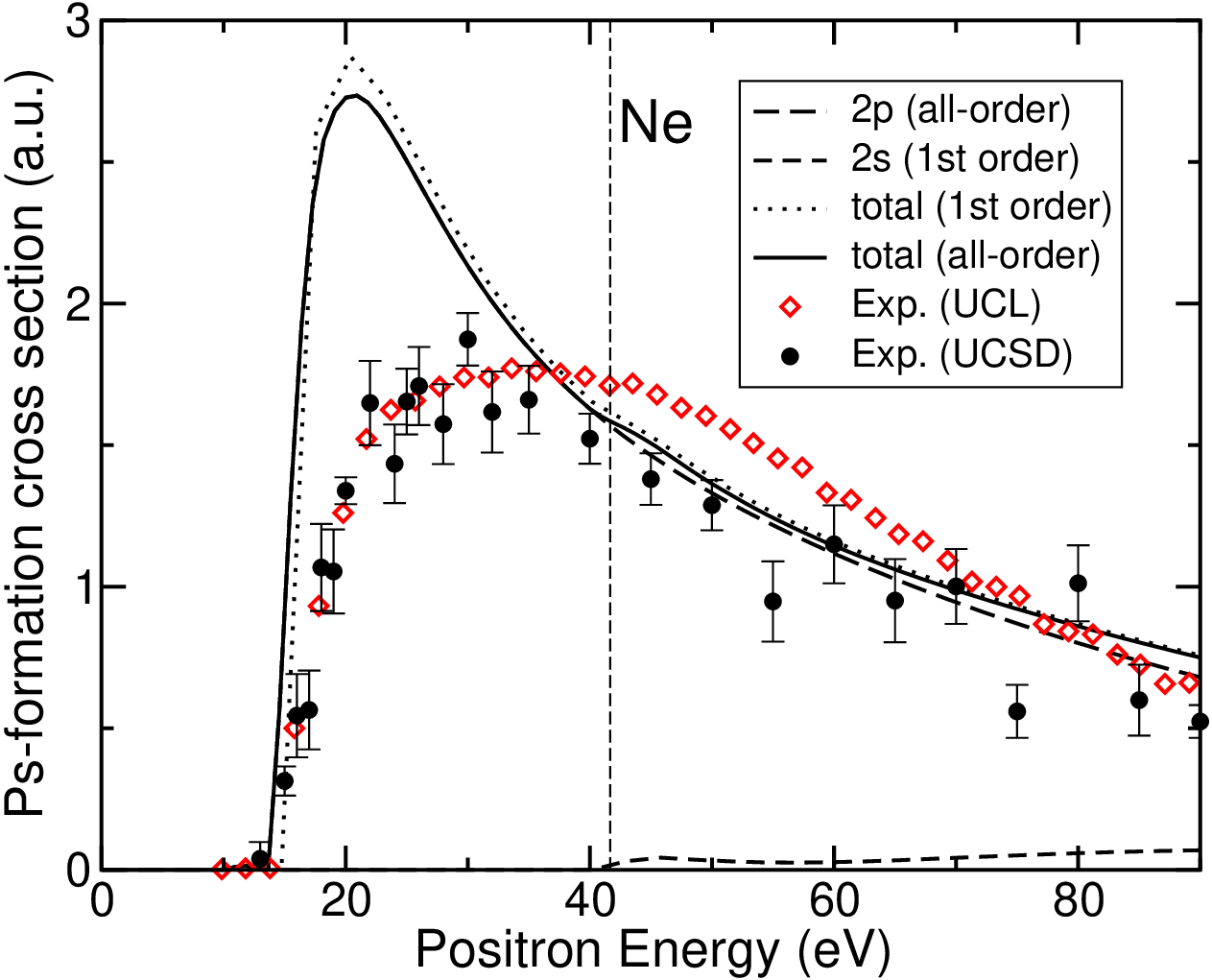}   
\end{minipage}
\begin{minipage}{6.8cm}
\includegraphics*[height=5cm]{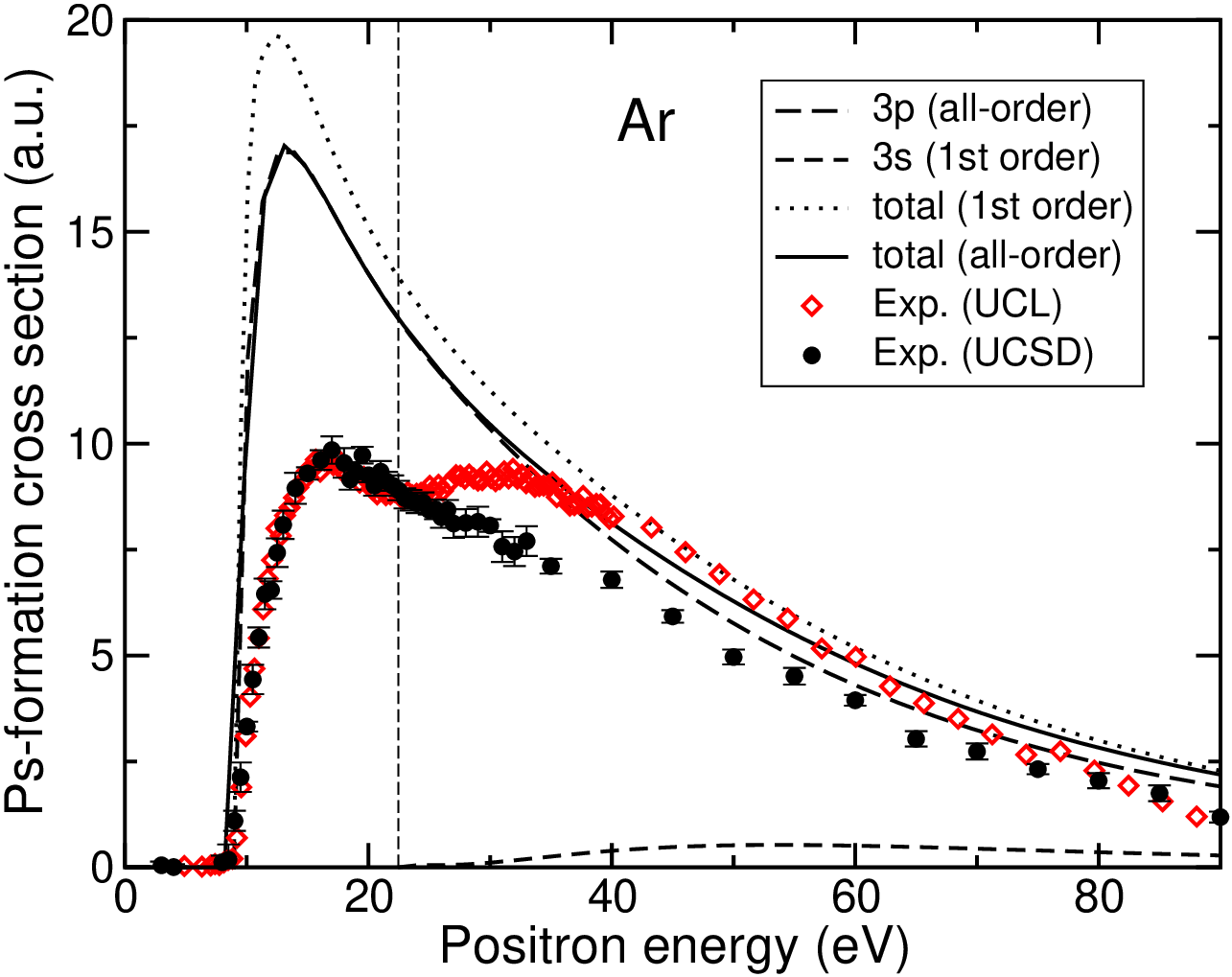}
\end{minipage}\\
\begin{minipage}{6.8cm}
\includegraphics*[height=5cm]{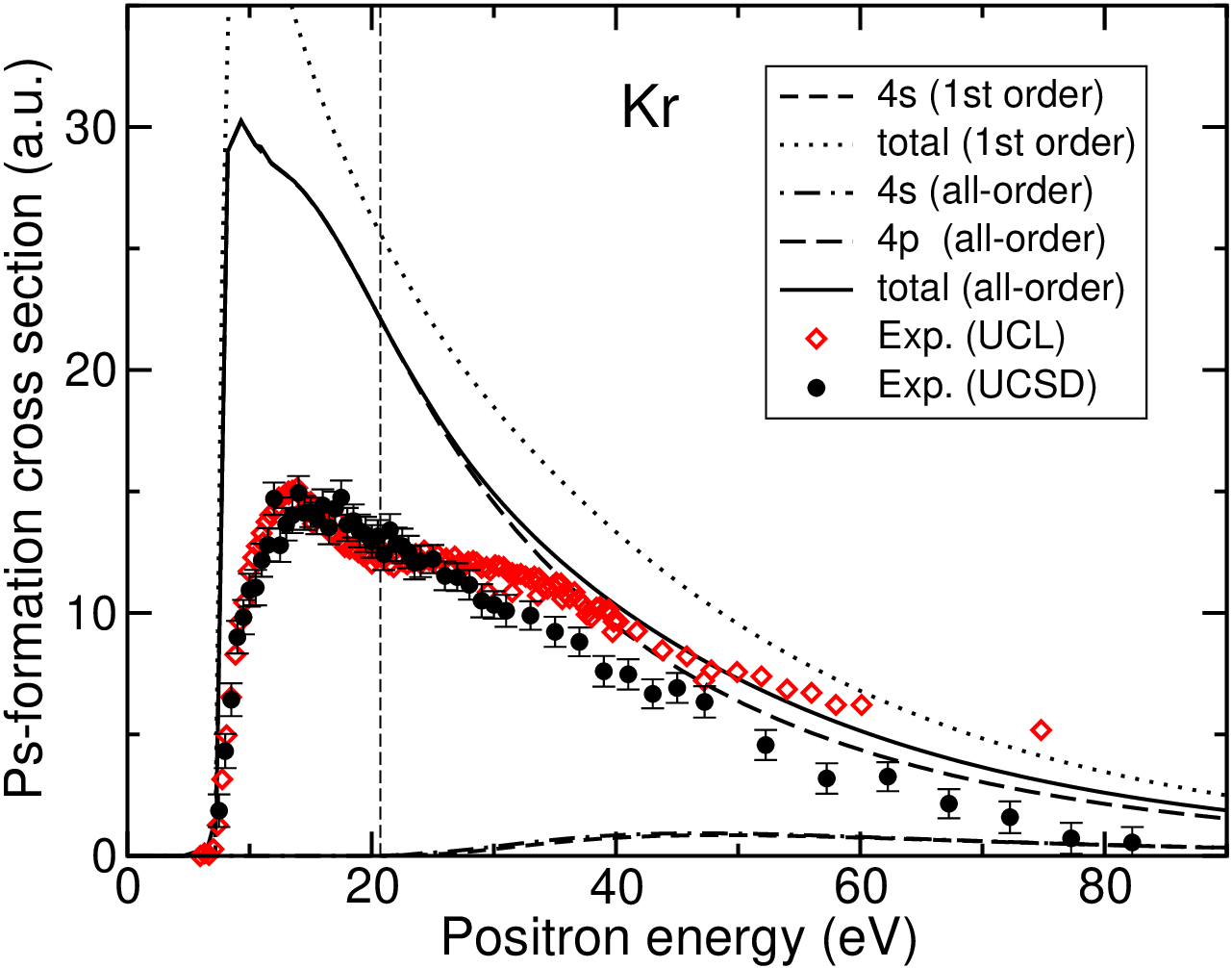}
\end{minipage}
\begin{minipage}{6.8cm}
\includegraphics*[height=5cm]{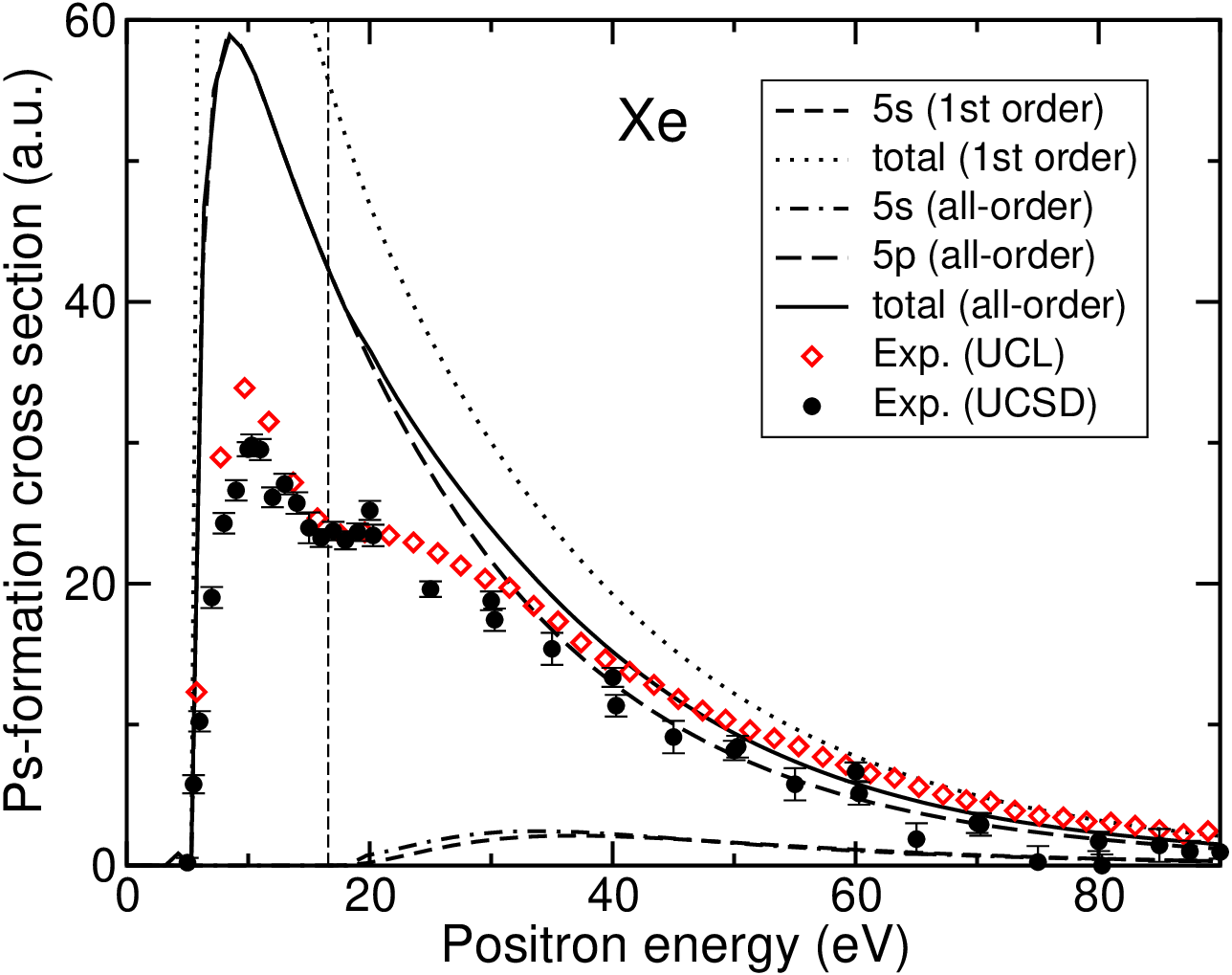}
\end{minipage}
\caption{Comparison of the calculated all-order and 1st-order Ps-formation
cross sections for Ne, Ar, Kr and Xe with experiment. Calculations:
dotted curve, total (1st-order); short-dashed curve, $ns$ subshell (1st order);
long-dashed curve, $np$ subshell (all-order); chain curve, $ns$ subshell
(all-order); solid curve, total cross section (all-order). Experiment:
solid circles, UCSD \cite{ma}; open diamonds, UCL \cite{la}. Vertical dashed
lines indicate Ps-formation thresholds for the subvalence $ns$ electrons.}
\label{fig:all}
\end{center}
\end{figure}

Figure \ref{fig:all} shows the all-order and 1st-order cross
sections for Ne, Ar, Kr and Xe along with the two sets of experimental data.
As expected, the difference between the all-order and 1st-order
results is greater for atoms where the Ps formation cross section is large,
i.e., for Kr and Xe. The all-order approximation has reduced the cross
section maxima for all atoms, but still not significantly enough to match
the experiment. However, theory and experiment are in
better agreement at higher positron energies. The situation looks especially
encouraging in Xe, where both experiments and theory are close above 40 eV.
Note that
for Kr and Xe, the all-order and 1st-order cross sections are markedly
different even at the higher-energy part of the scale. This emphasizes the
need for nonperturbative treatment of Ps formation.

Regarding the contribution of the subvalence $ns$ subshell, the corresponding
Ps-formation thresholds for Ne, Ar, Kr and Xe are 41.67, 22.47, 20.71 and
16.59 eV, respectively\footnote{Note that the $ns$ thresholds indicated
in figures 5--8 of Ref. \cite{la} are $ns$ {\em ionization} thresholds,
rather than the Ps-formation thresholds.}. The all-order
calculation was not performed for Ne and Ar, where we expected its effect
to be insignificant. In fact, the $2s$-subshell contribution to the
Ps-formation cross section in Ne is very small. Experimental data in Ne
also do not reveal any clear features that could be related to the opening
of a new channel at $\eps \approx 42$ eV.
The Ps formation from the $3s$ subshell in Ar is relatively more important.
The opening of this channel at about 22.5 eV coincides with the onset of the
second peak in the UCL data \cite{la}. It also marks the start of
a weak shoulder-like structure in the UCSD data \cite{ma}, where earlier
experiments \cite{CCG83} showed a much more prominent feature. The reason for
the discrepancy between different experimental observations is at present
unclear, but we suggest that the origin of this structure is most likely
related to the opening of the $3s$ Ps-formation channel.

Using the all-order approach makes a much greater difference in krypton and
xenon.
The valence $np$ contribution is greatly reduced in the energy range around
the maximum. In contrast, the subvalence $ns$ contribution is increased in
the all-order approach. The onset of the Ps-formation cross section
from the $4s$ orbital in Kr is very smooth. The UCSD data for Kr do not show
any feature just above 20 eV, while the UCL data possess a clear shoulder,
whose onset is more rapid than that predicted by the theory.
At higher energies the two sets of experimental data diverge, with our
all-order results being in-between. Given that the theory is
expected to be more reliable here than at lower energies, we suggest that
this discrepancy points to a need for further experimental studies.

Of the four atoms examined, Xe has the largest Ps formation cross
section by the subvalence electrons. Its contribution results in a change of
slope of the calculated total cross section. For Xe both sets of experimental
data show a clear shoulder-like structure, whose onset is close to the $5s$
Ps-formation threshold. The overall size of the shoulder is comparable with
the calculated $5s$ Ps formation cross section (chain curve in Figure
\ref{fig:all}). At higher energies there is good agreement between the two
experiments and the calculated cross section.

One may speculate that in a better calculation, the Ps-formation cross section
from the valence $np$ orbital will be suppressed around its maximum and
at the energies below 40 eV. Then, even if the $ns$ Ps-formation cross
sections remain close to the present estimates, their contribution to the
total will be more noticeable.

Finally, an alternative explanation of the shoulder/secondary-peak
structure discussed, e.g., in Ref. \cite{la}, is that it is caused by Ps
formation in excited states. According to the DWBA calculations of Gilmore
{\em et al.} \cite{gi}, this contribution is not negligible (though small).
However, the thresholds for excited-state Ps formation lie much lower than
the energies where the structures are observed, making its importance for
these structures questionable.

\section{Conclusions}

A comparison of the 1st-order and all-order results shows that as Ps formation
is strong, it cannot be treated perturbatively. Going beyond the 1st-order
approximation reduces the cross sections, especially at low energies. However,
below 40 eV the calculated cross sections are still higher than experimental
values.  Above this energy theory and experiment generally converge.
In particular, in Xe we observe good agreement between the calculations and
experimental data from the UCL and UCSD groups. 

There are two reasons for the discrepancy between theory and experiment
at low energies. First, the motion of Ps in our calculations is described 
by a plane wave. The electron part of the Ps wavefunction is orthogonalized
to the target electron orbitals. This manifestation of the Pauli principle
to some extent accounts for the interaction between the Ps and the
final-state ion. On the other hand, the positron repulsion from the nucleus
in the final Ps state is completely neglected. This repulsion is
especially important for the lower positron partial waves. Its neglect
is probably the main reason for the overestimation of the Ps-formation cross
section maxima by the present method.

Secondly, all open channels, i.e., elastic scattering, Ps formation and
direct ionization must be included simultaneously. Above the atomic
ionization threshold all of these channels compete for the
positron flux. This effect can be accounted for by the correlation potential
method described in Sec. \ref{subsec:all}, by adding the lowest 2nd-order
contribution to $\Sigma _E$ (see, e.g., \cite{GK96}). However, to be able
to extract the Ps formation cross section from such a calculation,
the formalism of Sec. \ref{subsec:all} must be extended. In the present
form the imaginary part of the phaseshift allows one to find only
the total reaction cross section.

One may expect that both of these effects will make the Ps formation cross
section smaller, and bring it into a close agreement with experiment.
By further including Ps formation from the inner valence subshell and Ps
formation in excited states, one should achieve a complete description of
the Ps formation process, including any secondary structures.

\section*{Acknowledgements}
We are grateful to G. Laricchia and D. Murtagh (UCL), C. Surko and J. Marler
(San Diego) and H. R. J. Walters and S. Gilmore (Belfast) for providing us
with their data in table form and useful discussions. LJMD acknowledges
DEL (Northern Ireland) for support in the form of a PhD studentship.


\begin{thebibliography}{10}\setlength{\itemsep}{-3pt}

\bibitem{la} G.~Laricchia, P.~Van~Reeth, M.~Sz\l ui\'nska and J.~Moxom
J. Phys. B {\bf 35} (2002) 2525.

\bibitem{ma} J.~P. Marler, J.~P.~Sullivan and C.~M.~Surko, Phys. Rev. A
{\bf 71} (2005) 022701.

\bibitem{gi} S.~Gilmore, J.~E.~Blackwood and H.~R.~J.~Walters,
Nucl. Instrum. Methods B {\bf 221} (2004) 129.

\bibitem{mca}M. T. McAlinden and H. R. J. Walters, Hyperfine
Interactions {\bf 73} (1992) 65.

\bibitem{CCG83}M. Charlton, G. Clark, T. C. Griffith, and G. R. Heyland,
J. Phys. B {\bf 16} (1983) L465.

\bibitem{oh} T. Ohdaira, R. Suzuki, Y. Kobayashi, T. Akahane and L. Dai,
Appl. Surface Science {\bf 194} (2002) 291.

\bibitem{DFG96}V.~A.~Dzuba, V.~V.~Flambaum, G.~F.~Gribakin and W.~A.~King,
J. Phys. B {\bf 29} (1996) 3151.

\bibitem{humberston} J.~W.~Humberston, Can. J. Phys. {\bf 60} (1982) 591.
 
\bibitem{vanreeth1} P.~Van Reeth, J.~W.~Humberston, J. Phys. B {\bf 28}
(1995) L511. 
 
\bibitem{vanreeth2} P.~Van Reeth, J.~W.~Humberston, J. Phys. {\bf 30}
(1997) L95.  
 
\bibitem{ward} S.~J.~Ward, J.~H.~Macek and S.~Yu.~Ovchinnikov,
Phys. Rev. A {\bf 59} (1999) 4418.
 
\bibitem{ACC82} M. Ya. Amusia, N. A. Cherepkov, L. V. Chernysheva, D. M. 
Davidovi\'c, and V. Radojevi\'c, Phys. Rev. A {\bf 25} (1982) 219.

\bibitem{GK96}G. F. Gribakin and W. A. King, Can. J. Phys. {\bf 74}
(1996) 449.

\end{thebibliography}
\end{document}